\documentclass[runningheads]{llncs}
\usepackage[T1]{fontenc}
\usepackage{graphicx}
\usepackage{amsmath}
\usepackage{amssymb}
\usepackage{booktabs}
\usepackage{tabularx}
\usepackage{multirow}
\usepackage{xcolor}
\usepackage{hyperref}
\usepackage[protrusion=true,expansion=false]{microtype}
\usepackage{enumitem}
\usepackage{tikz}
\usepackage{pgfplots}
\pgfplotsset{compat=1.18}
\usetikzlibrary{arrows.meta, shapes.geometric, positioning, fit, backgrounds, calc}
\usepackage{array}
\usepackage{longtable}
\usepackage{caption}
\usepackage{pgfplotstable}
\usetikzlibrary{patterns}

\definecolor{gemblue}{RGB}{58,111,160}
\definecolor{clared}{RGB}{176,58,46}
\definecolor{pitline}{RGB}{192,57,43}
 
\definecolor{hdrblue}{RGB}{6,90,130}
\definecolor{rowalt}{RGB}{234,244,251}
\definecolor{pitgreen}{RGB}{30,132,73}
\definecolor{pityellow}{RGB}{232,131,10}
\definecolor{pitred}{RGB}{192,57,43}
 
\hypersetup{
  colorlinks=true,
  linkcolor=hdrblue,
  citecolor=hdrblue,
  urlcolor=hdrblue,
  hidelinks
}
\usepackage{pifont}

\usepackage{graphicx}
\begin{document}
\title{When Agents Control Robots: A Zero Trust Policy Model for Agentic Cyber-Physical Systems}
%
%
\author{Tharindu Ranathunga\inst{1} \and
Kavishka Fernando\inst{1} \and
Susan Rea\inst{1}}

\institute{Nimbus Centre, Munster Technological University, Cork, Ireland\\
\email{tharindu.ranathunga@mtu.ie}}
%
%
%
\maketitle              

\begin{abstract}
Multi-agent systems powered by large foundation models (LFMs) are
increasingly deployed to control industrial robots through natural
language, creating deployments in which security failures produce
physical consequences. We analyse this threat landscape through
Cobot-Claw, a deployed four-agent system for UR3e robotic arm
control, and identify five attack classes specific to agentic
cyber-physical systems. We propose ZTPM, a Zero Trust Policy Model comprising 25 typed primitives across five enforcement domains with Physical Impact Tiers as a runtime policy dimension. An empirical evaluation across 60 execution traces on two LFM backends provides initial evidence that actuation parameter selection is model-dependent and non-deterministic, motivating the need for policy-level enforcement at the physical actuation boundary.
\vspace{-5mm}
\end{abstract}


\section{Introduction}

Autonomous agents powered by large foundation models
(LFMs), encompassing large language models (LLMs),
vision-language models (VLMs), and vision-language-action
models (VLAs)~\cite{brohan2023rt2}, are increasingly
deployed in systems that interact with the physical world.
Unlike earlier AI systems, LFM-based agents reason over
context, invoke external tools through standardised
interfaces such as the Model Context Protocol
(MCP)~\cite{mcp2025}, coordinate sub-agents through
delegation chains, and maintain a persistent state across
extended task horizons~\cite{atf2026,ms2026}.

When LFM-based agents are deployed to control
cyber-physical infrastructure, a qualitatively new class
of system emerges. We use the term \emph{Multi-Agentic
Cyber-Physical Systems} (MA-CPS) to describe deployments
in which agents perceive physical environments through
sensor pipelines, reason over that data, and issue
commands to actuators through tool interfaces.
Autonomous agents controlling industrial robots are one
instance of this class; others include agents managing
building automation, autonomous vehicles, or chemical
process control. What distinguishes an MA-CPS from both a
traditional cyber-physical system and a digital-only
agentic system is the closed physical feedback loop: sensor
data enters agent reasoning, reasoning produces actuation
commands, actuation changes the physical world, and the
changed world generates new sensor data.


This architecture introduces security challenges that
existing frameworks were not designed to address.
Adversarial instructions can propagate through delegation
chains until they reach agents capable of physical
actuation. Perception agents can be manipulated through
sensor inputs alone, causing downstream agents to plan
motions against a false workspace state. Memory and
retrieval systems introduce persistent attack surfaces
through poisoned context. Existing policy enforcement
remains fundamentally stateless, unable to detect unsafe
physical outcomes from sequences of individually compliant
actions. As discussed in Section~\ref{sec:related},
frameworks for agentic AI
security~\cite{owasp2025,atf2026,maestro2025}, industrial
control standards~\cite{iec62443,nist80082}, embodied AI
security research~\cite{xing2025embodied}, and Zero Trust
Architecture~\cite{nist800207} each address part of this
landscape, but none governs the full path from LFM
reasoning to physical actuation.

This paper makes three contributions:
\begin{enumerate}[label=\arabic*.]
  \item A formal MA-CPS system model and a taxonomy of
    five attack classes for agentic robotic environments.
  \item A Zero Trust Policy Model for MA-CPS comprising
    25 typed policy primitives across five enforcement
    layers with runtime Physical Impact Tiers.
  \item A deployment-grounded evaluation on
    \emph{Cobot-Claw}, a four-agent UR3e robotic control
    system, including empirical evidence from 60 traces
    across two LFM backends and coverage analysis
    across all five attack classes.
\end{enumerate}
\section{Related Work}
\label{sec:related}

Zero Trust Architecture (ZTA) was formalised by NIST in SP
800-207~\cite{nist800207}, establishing continuous verification
of every access request against policy. Policy languages including XACML~\cite{xacml2013}, OPA with Rego~\cite{opa2023}, and Cedar~\cite{cedar2024} provide expressive attribute-based access control for digital resources.
However, all operate on a request/response model scoped to digital
objects: none defines policy primitives for physical actuation,
sensor data provenance, or tool invocation sequences with
consequence-tiered enforcement.

The security of agentic AI systems has received growing attention.
The OWASP Top~10 for Agentic AI~\cite{owasp2025} catalogues key
risks including prompt injection, excessive agency, and memory
poisoning. The Agentic Trust Framework~\cite{atf2026} addresses
agent identity and delegation-chain governance.
MAESTRO~\cite{maestro2025} provides a seven-layer threat model
for agentic AI. Red Hat~\cite{redhat2026} addresses the delegation
token exchange problem in agent-to-agent communication.
Microsoft~\cite{ms2026} extends ZTA to the full AI lifecycle.
Rebedea et al.~\cite{rebedea2023nemo} present NeMo Guardrails, a
runtime toolkit for adding programmable rails to LLM conversational
output; it operates on single-agent responses and does not address
multi-agent delegation or physical tool invocation. These frameworks collectively advance governance of agentic AI in digital environments but share a structural limitation: all model the worst-case consequence of a policy violation as a data breach or unauthorised API call. None addresses physical actuation as a
policy object class or provides consequence-tiered enforcement.

Industrial CPS security is governed by IEC~62443~\cite{iec62443}
and NIST SP~800-82~\cite{nist80082}, which define security levels
and access control for operational technology environments.
Robot safety and middleware security frameworks, including
ISO/TS~15066~\cite{iso15066} and ROS/ROS~2 security mechanisms,
provide important constraints such as communication protection,
access control, speed limits, force limits, and workspace safety
rules. However, these approaches primarily assume deterministic
control logic or middleware-level enforcement; they do not govern
how LFM-based agents interpret context, propagate delegated
authority, or transform natural-language goals into physical
command sequences. Xing et al.~\cite{xing2025embodied} survey
vulnerabilities in embodied AI, categorising exogenous threats such
as sensor spoofing and adversarial patches, but do not address the
consequence chain that emerges when perception is mediated by an
LFM reasoning layer. Vision-language-action models such as
RT-2~\cite{brohan2023rt2} demonstrate the frontier of
LFM-mediated robotic control. The Model Context
Protocol~\cite{mcp2025} provides a standardised tool invocation
interface now bridging LFM agents to physical hardware. These
standards, safety mechanisms, and protocols do not model
LFM-based agents as control-plane components whose delegated
decisions require runtime governance at the physical actuation
boundary.

No existing framework simultaneously addresses agentic cognitive
operations as policy subjects, physical actuation as a policy
object class with consequence tiers, multi-agent delegation chains
as trust propagation paths, and runtime policy enforcement for
LFM-driven physical actuation. ZTPM is designed to address this
gap.
\section{System Model and Security Analysis}
\label{sec:system}

\subsection{Formal System Model}
\label{subsec:sysmodel}

We focus on agentic robotic control systems in which
LFM-based agents interpret operator goals, reason over
sensor-derived state, and issue commands to industrial
robot arms, mobile platforms, or end-effector tools.
We model this class as a Multi-Agentic Cyber-Physical
System (MA-CPS):
\begin{equation}
  \mathrm{MA\mbox{-}CPS} =
  \langle A,\; O,\; E,\; T,\; \Phi,\; H \rangle .
  \label{eq:macps}
\end{equation}

Here $A = \{a_1,\ldots,a_n\}$ is a set of LFM-based
agents, including orchestrators and specialist agents for
robotic execution, perception, planning, configuration, or
safety monitoring. Each agent maintains task context,
memory, and tool interfaces, and may delegate to other
agents, forming delegation chains of depth $d \geq 1$.

$O = O_d \cup O_p$ partitions policy objects into digital
objects $O_d$ and physical objects $O_p$. In robotic
systems, $O_d$ includes prompts, memory stores, RAG
knowledge bases, task plans, tool schemas, and robot
controller interfaces, while $O_p$ includes manipulators,
mobile bases, end-effectors, sensors, workpieces, humans in
the workspace, and the physical operating environment. This
digital--physical partition is the key CPS extension.

\begin{figure}[t]
\centering
\includegraphics[width=\columnwidth]{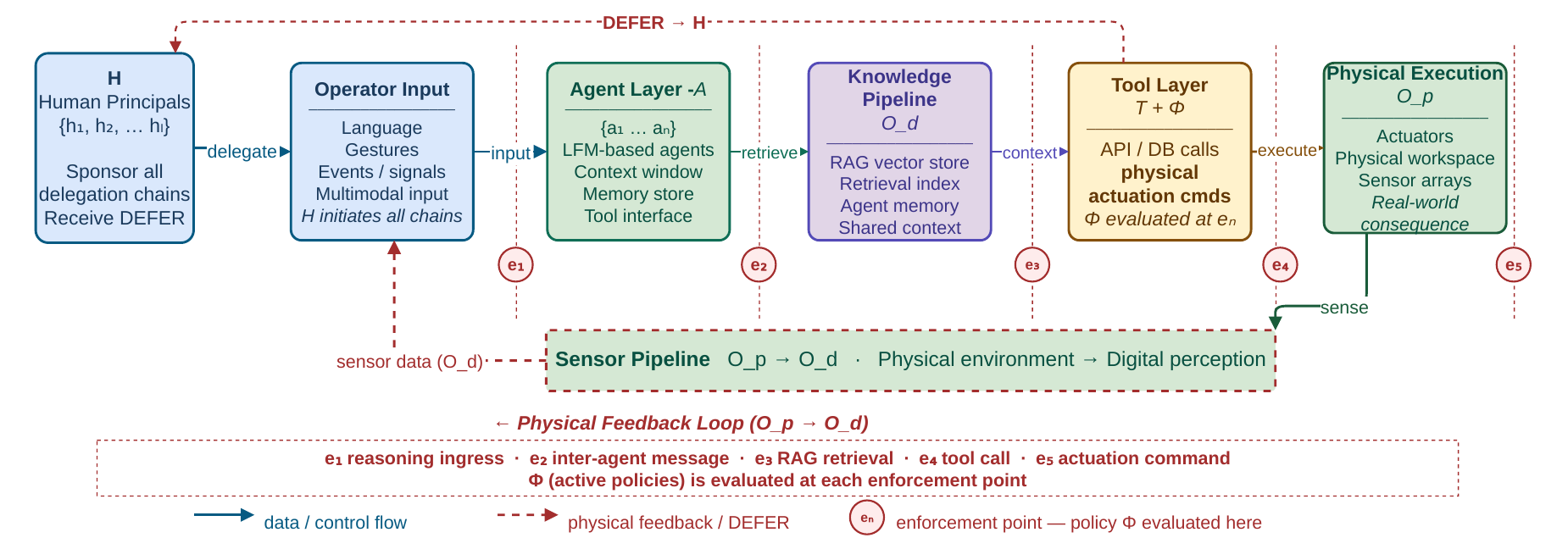}
\caption{Generic MA-CPS architecture for agentic robotic
control, showing enforcement points $e_1$--$e_5$ and the
physical feedback loop.}
\label{fig:macps}
\vspace{-4mm}
\end{figure}

$E$ is the set of enforcement points, including reasoning
ingress, inter-agent message boundaries, RAG retrieval,
memory writes, tool calls, and pre-actuation command
boundaries. $T$ is the agent tool set; tools that can change
$O_p$ are \emph{physical tools} and carry a Physical Impact
Tier classification. $\Phi$ is the active policy set,
defined in Section~\ref{sec:model}. $H$ is the set of human
principals who sponsor delegation chains and receive DEFER
decisions. The sensor pipeline closes the physical feedback
loop from actuation to perception. Figure~\ref{fig:macps}
illustrates the architecture.

\subsection{Adversary Model}
\label{subsec:threat}

We assume an adversary capable of injecting content through
any input channel, including operator inputs, sensor streams,
RAG stores, tool server responses, and inter-agent messages.
The adversary cannot modify LFM weights, cryptographic
material, or hardware firmware. The adversary's objectives
are to induce unsafe physical actions, exfiltrate
operational data, disrupt system availability, or establish
persistent influence through manipulation of knowledge
stores.

A property unique to MA-CPS is \emph{delegation-chain
amplification}: an injection received by an orchestrator or
upstream agent may propagate to downstream agents with more
consequential tool access. The physical consequence of an
adversarial input is therefore bounded not by the privilege
of the agent that first receives it, but by the most capable
agent reached through the delegation chain.

\subsection{Attack Taxonomy}
\label{subsec:attacks}

We identify five attack classes that arise in agentic
robotic MA-CPS when LFM reasoning, multi-agent delegation,
sensor-driven perception, and physical actuation are
combined. Existing frameworks address some underlying
mechanisms, such as prompt injection or tool misuse, but not
their consequence chain to robotic actuation.



\textbf{AC-1: Adversarial Perception Injection.}
An adversary manipulates the physical environment within
sensor range to corrupt data at the physical-to-digital
conversion boundary, causing perception agents to construct
a false world model from which unsafe action plans are
generated. In a robotic workcell, this may involve
manipulating objects, markers, or occlusions within the
camera field so that the perception agent grounds a task
against a false workspace state. Physical sensor
manipulation is documented in IoT and autonomous
systems~\cite{nist80082}; what distinguishes AC-1 in
MA-CPS is the consequence chain: corrupted sensor data
enters LFM reasoning, is grounded against a knowledge base,
and cascades through a delegation chain to physical
actuators.

\textbf{AC-2:  Cross-Agent Prompt Propagation.}
A malicious instruction propagates through the delegation
chain without sanitisation at inter-agent boundaries,
reaching an agent with physical tool access. Prompt
injection is well documented~\cite{owasp2025}; however,
its propagation to physical actuators through delegation
chains is not addressed by existing frameworks.

\textbf{AC-3:  CPS Context Poisoning.}
An adversary corrupts agent reasoning through active prompt
injection, cross-session memory poisoning, or RAG store
poisoning. Each path produces the same outcome: the agent
reasons from false premises and generates physically unsafe
action plans. The poisoning mechanisms are catalogued in
MITRE ATLAS~\cite{mitreatlas2024} under techniques
including training data poisoning and prompt
injection~\cite{owasp2025,maestro2025}; their consequence
chain to unsafe physical actuation is not addressed by
existing frameworks.

\textbf{AC-4: Tool Scope Escalation.}
An agent acquires tool permissions beyond its delegated
scope by manipulating configuration components or spawning
sub-agents with broader authority. In robotic systems, such
escalation may grant access to motion, force, speed, or
end-effector commands outside the agent's delegated
authority. MITRE ATLAS catalogues related techniques under
AML.T0053 and AML.T0085.001~\cite{mitreatlas2024}; the
physical consequence of escalation in MA-CPS is the ability
to command hardware beyond authorised parameters.

\textbf{AC-5:  Physical Actuation Sequence Abuse.}
A sequence of individually compliant tool calls produces a
combined physical effect that is dangerous, visible only at
the sequence level. In an industrial robot arm, individually
valid joint-space or Cartesian moves may combine into an
unsafe trajectory, collision risk, or workspace boundary
violation. Existing tool governance frameworks evaluate
invocations independently; they do not evaluate the combined
physical effect of an action sequence.

Section~\ref{sec:evaluation} instantiates the model and
attack taxonomy on a deployed multi-agent robotic control
system, demonstrating coverage across all five classes.
\section{Zero Trust Policy Enforcement for Agentic Robotic Systems}
\label{sec:model}

Zero Trust Architecture establishes that no entity should be trusted by default and that every access request must be continuously verified. In agentic robotic systems, this principle must extend beyond network flows and identities to cover agent delegation, cognitive inputs, tool invocation, inter-agent communication, and physical actuation. We propose
ZTPM for MA-CPS, a Zero Trust Policy Model for LFM-mediated robotic control. The model governs the path from operator input to physical execution through five enforcement layers and uses Physical Impact Tiers (PIT) to escalate enforcement according to runtime physical risk.

\subsection{Enforcement Layers}
\label{subsec:layers}

ZTPM defines five enforcement layers ordered from operator input to robotic execution. Each layer corresponds to a trust boundary in an agentic robotic control pipeline: agent identity and delegation, cognitive input integrity, tool execution authority, cross-agent trust propagation, and adaptive behavioural governance. Layers L1 through L4 govern individual requests at the enforcement points defined
in Section~\ref{sec:system}. L5 operates over time: it
monitors behavioural patterns across a window of
interactions and intervenes when agent behaviour deviates
from a policy-conformant baseline. These layers correspond to the policy domains shown in Figure~\ref{fig:primitives}.


\subsection{Physical Impact Tiers}
\label{subsec:pit}

Robotic actions do not carry uniform risk. A motion command that is safe in an empty workspace may become safety-critical when a human, fragile workpiece, or restricted zone is detected nearby. Physical Impact Tiers (PIT) capture this runtime risk by classifying each tool invocation before it is executed and selecting the corresponding enforcement response.

Every tool in $T$ carries a base PIT assigned at registration time. At runtime, this base class may be elevated by the invocation parameters and by the live physical context reported by sensors or perception agents. The runtime PIT is:

\begin{equation}
  \mathit{pit}(\mathit{inv}) = \max\!\big(
    \mathit{tool.pit\_class},\;
    \mathit{param\_pit}(\mathit{inv.params}),\;
    \mathit{context\_pit}(\mathit{env})
  \big)
  \label{eq:pit}
\end{equation}

For example, a low-speed gripper action may remain PIT-1 in an isolated workspace, while a robot-arm movement near a human operator may be elevated to PIT-3 and require DEFER. This runtime, sensor-informed escalation is the key CPS-specific property of PIT: the same command may follow different enforcement paths depending on the live robotic workspace state. Table~\ref{tab:pit} defines the five tiers.

\begin{table}[t]
\caption{Physical Impact Tier classification and enforcement.}
\label{tab:pit}
\small
\renewcommand{\arraystretch}{1.15}
\begin{tabularx}{\textwidth}{@{}lp{2.0cm}>{\raggedright\arraybackslash}X>{\raggedright\arraybackslash}p{2.4cm}@{}}
\toprule
\textbf{Tier} & \textbf{Name} & \textbf{Definition} &
  \textbf{Enforcement} \\
\midrule
PIT-0 & No effect &
  Digital only; no physical consequence &
  PERMIT + AUDIT \\
PIT-1 & Reversible &
  Fully reversible within normal parameters &
  PERMIT + AUDIT \\
PIT-2 & Consequential &
  Reversible with effort; minor risk if incorrect &
  PERMIT if trust $\geq\theta$; else DEFER \\
PIT-3 & High-consequence &
  Equipment damage or minor harm if incorrect &
  DEFER; DENY on timeout \\
PIT-4 & Safety-critical &
  Potential serious injury or infrastructure damage &
  DENY; dual authorisation only \\
\bottomrule
\end{tabularx}
\vspace{-3mm}
\end{table}

\subsection{Policy Model}
\label{subsec:policymodel}

A policy in ZTPM is a 7-tuple:

\begin{equation}
  \pi = \langle \mathit{Subject},\; \mathit{Object},\;
  \mathit{Predicate},\; \mathit{EP},\; \mathit{Effect},\;
  \mathit{Obligation},\; \mathit{PITBound} \rangle .
  \label{eq:policy}
\end{equation}

$\mathit{Subject} \in (A \cup H)$ is the agent or human
principal whose action is governed. $\mathit{Object} \in
(O_d \cup O_p)$ is the governed resource or operation; the inclusion of $O_p$ allows policies to refer directly to robotic actuation, sensors, tools, and workspace state. $\mathit{Predicate}$ is an evaluable condition over request context at enforcement point $\mathit{EP} \in E$. $\mathit{Effect} \in \{\mathit{PERMIT}, \mathit{DENY}, \mathit{DEFER}\}$, where DEFER suspends the action and routes it to a human principal $h \in H$. $\mathit{Obligation}$ captures mandatory side-effects such as audit logging, telemetry, and trust-score updates. $\mathit{PITBound}$ defines the minimum physical impact tier at which the policy
escalates from PERMIT to DEFER or DENY; $\varnothing$
indicates no physical consequence dependency.

\begin{figure}[t]
\centering
\includegraphics[width=1\columnwidth]{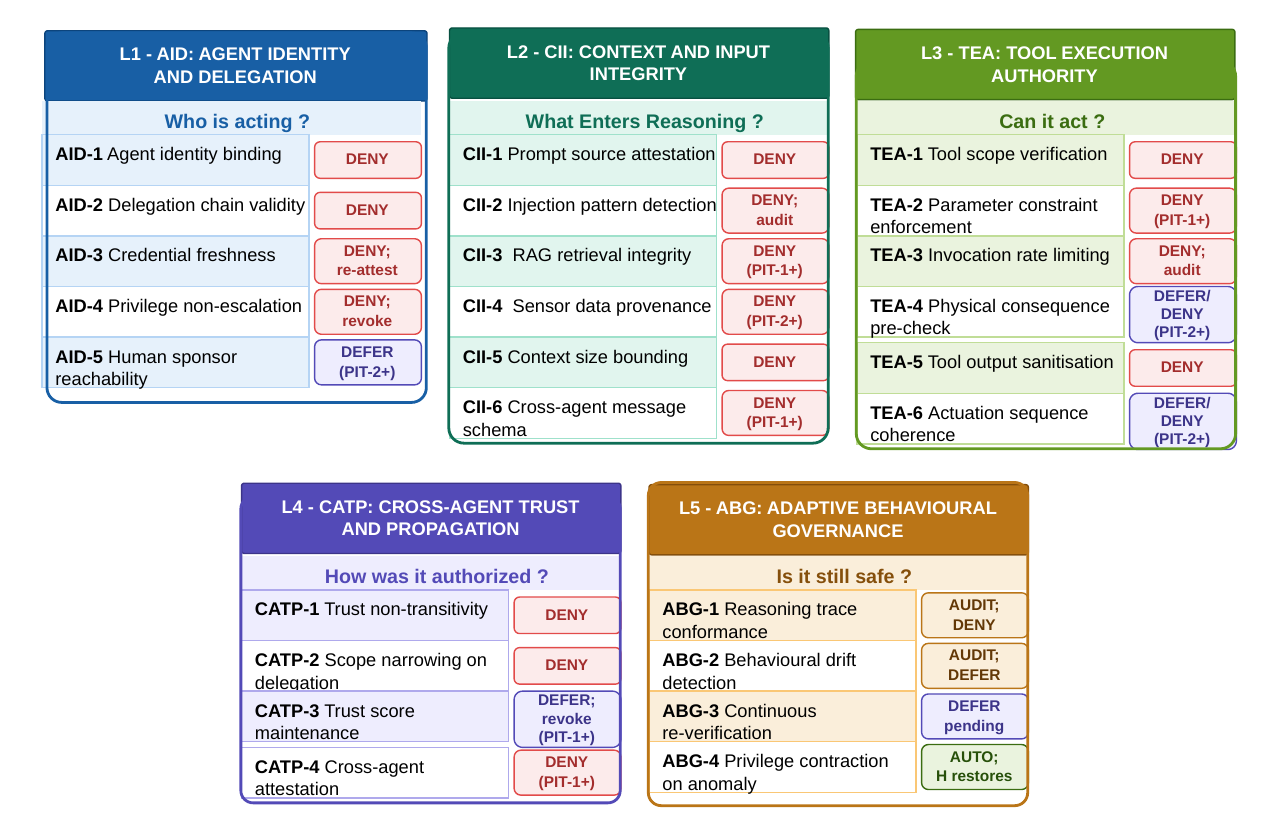}
\caption{ZTPM policy primitives and enforcement outcomes.}
\label{fig:primitives}
\vspace{-4mm}
\end{figure}

When the runtime PIT of an invocation reaches or exceeds
$\pi.\mathit{PITBound}$, the effect escalates. PIT-3
requires human approval through DEFER, while PIT-4 is denied unless prior dual authorisation has been granted. This makes physical consequences a first-class input to policy evaluation rather than an external safety check. ZTPM organises policies into five domains across the enforcement layers, comprising 25 typed primitives. Figure~\ref{fig:primitives} lists the primitives and their enforcement outcomes.

\textbf{D1: Agent Identity and Delegation (AID).}
AID answers \emph{WHO is acting?} by verifying agent
identity and delegation-chain validity before any downstream action is considered. Every delegation chain must be rooted in $H$, must not escalate scope, and must contain only valid, non-revoked links. For robotic systems, this prevents a planning or configuration agent from indirectly acquiring
motion or end-effector authority outside its delegated role.

\textbf{D2: Cognitive Input Integrity (CII).}
CII answers \emph{WHAT enters reasoning?} by validating
operator inputs, sensor data, RAG-retrieved content,
inter-agent messages, and tool responses before they enter agent context. In robotic control, this includes provenance checks for perception data and integrity checks for manuals, scenario libraries, or task memories used to ground physical action plans.

\textbf{D3: Tool Execution Authority (TEA).}
TEA answers \emph{CAN it act?} by governing every tool
invocation before execution. It verifies that the requested robot API, motion command, gripper action, scenario execution, or configuration change is within the agent's
delegated scope and parameter limits. TEA also performs the physical consequence pre-check and sequence-level coherence checks required before robotic actuation.

\textbf{D4: Cross-Agent Trust Propagation (CATP).}
CATP answers \emph{HOW was it authorised?} by ensuring that trust is explicit and non-transitive across agent boundaries. Delegation can only narrow scope, each boundary requires re-attestation, and dynamic trust scores are updated using interaction history, anomaly signals, and time decay. This prevents trust granted to an orchestrator from automatically flowing to every downstream robotic or perception agent.

\textbf{D5: Adaptive Behavioural Governance (ABG).}
ABG answers \emph{IS IT STILL SAFE?} by monitoring agent
behaviour across time rather than evaluating only individual actions. When behaviour deviates from the policy-conformant baseline, ABG triggers re-verification and contracts the agent's scope to the minimum safe set until a human principal restores it. In robotic systems, this captures gradual drift such as repeated near-boundary movements, unusual tool-use patterns, or escalating command sequences that remain safe only when viewed in isolation.
\section{Evaluation \& Discussion}
\label{sec:evaluation}

\subsection{System Instantiation}
\label{sec:instantiation}

We evaluate ZTPM by instantiating it on \emph{Cobot-Claw},
a deployed four-agent robotic control system for natural
language operation of a Universal Robots (UR) UR3e industrial robotic arm. This evaluation is a first step toward empirical validation:
in this work, we ground ZTPM in a real robotic deployment,
map the formal MA-CPS model to the implementation
architecture, identify enforcement points in the control
pipeline, and analytically assess coverage of the five attacks
classes defined in Section~\ref{subsec:attacks}.

\begin{figure}[t]
\centering
\includegraphics[width=\columnwidth]{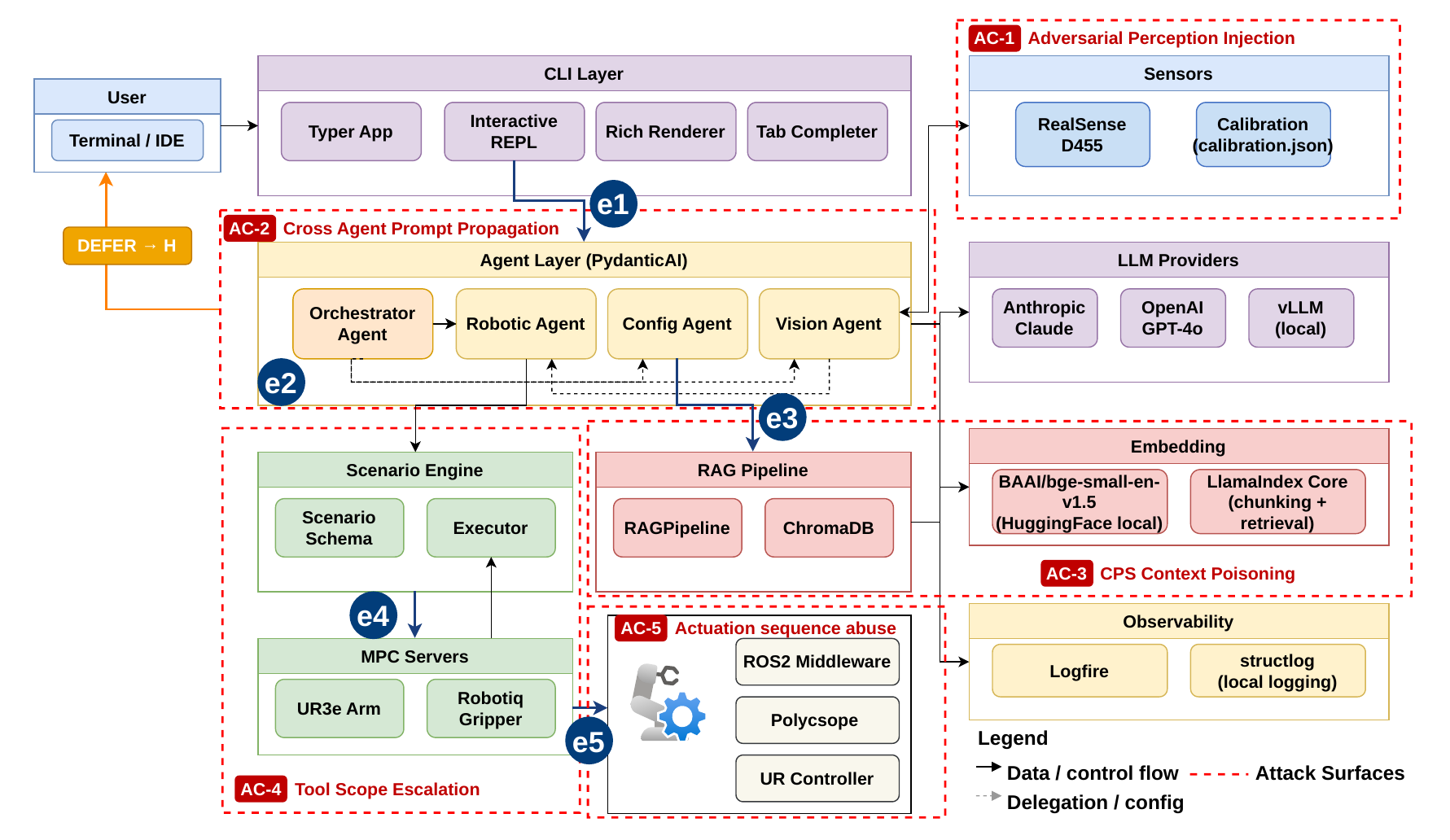}
\caption{Case study: Cobot-Claw.}
\label{fig:cobot-claw}
\vspace{-4mm}
\end{figure}

Figure~\ref{fig:cobot-claw} shows the cobot-claw
architecture and overlays the ZTPM enforcement points and
attack surfaces. $H$ is the human operator interacting
through a CLI, which accepts commands in natural language. $A$ comprises the Orchestrator, Robotic, Vision, and Config agents, each
implemented using PydanticAI and connected to local or cloud
LLM and VLM backends. $O_d$ includes the ChromaDB RAG store, which comprises the UR-specific knowledge base, scenario schema, and agent context windows. $O_p$ includes the UR3e arm, Robotiq
gripper, workspace, and a 3D vision sensor field. $T$ is implemented using MCP, which comprises tools to do robotic arm movements and gripper actions. 

The enforcement points are instantiated as follows: $e_1$ is the
CLI-to-Orchestrator boundary; $e_2$ is the inter-agent
delegation boundary; $e_3$ is the context admission boundary
for RAG, memory, scenario, and perception-derived context;
$e_4$ is the MCP tool invocation boundary; and $e_5$ is the
pre-actuation boundary before commands reach the UR controller. The physical feedback loop closes from the UR workspace through the 3D vision sensor pipeline to the Vision Agent context window, matching the $O_p \to O_d$ loop in Section~\ref{sec:system}. The Logfire observability stack instruments these boundaries, providing
the monitoring surface required by the ABG layer and a path
toward empirical evaluation.

\subsection{Need for Runtime PIT Enforcement}
\label{sec:pit-enforcement-eval}

The system instantiation above identifies $e_5$ as the
pre-actuation boundary before commands reach the UR controller.
This boundary is the point at which PIT enforcement must be
applied if physical consequence is to be governed independently
of LFM behaviour. We therefore tested whether safety-relevant
actuation parameters are reliably modulated by the LFM alone,
or whether they require policy-level enforcement.

We ran a controlled experiment on the deployed Cobot-Claw system
using the same neutral motion task across three workspace risk
conditions. The task asked the robot to execute a sweeping arc
across the workspace and contained no speed-related language.
The three conditions matched the PIT tiers used in the policy
model: C0, an empty workspace corresponding to PIT-1; C1, a
fragile object placed on the table below the arm path,
corresponding to PIT-2; and C2, a human operator positioned
0.5\,m from the robot base, corresponding to PIT-3. Each
condition was executed 10 times with each of the two LFM
backends available in the Cobot-Claw deployment: Gemma~4,
running locally via vLLM, and Claude Sonnet~4.6, accessed
through an API. This produced 60 execution traces. For each
trace, we computed the mean actuation speed across all physical
MCP tool calls. The resulting distributions are shown in
Figure~\ref{fig:eval} (left).

The results reveal two distinct failure modes. Gemma~4 produced
nearly identical speeds in all three conditions, centred at
0.500\,rad/s with SD\,$<$\,0.001. The model did not adapt
actuation speed to workspace risk: the presence of a human
operator produced the same physical behaviour as an empty
workspace.

Claude Sonnet~4.6 showed partial context sensitivity: mean
speed fell from 0.42\,rad/s under C0 to 0.17 under C1 and
0.21 under C2. However, the risk ordering was inconsistent
with the PIT classification. The fragile-object condition
(C1, PIT-2) produced lower speeds than the human-operator
condition (C2, PIT-3), even though C2 carries the higher
physical impact tier. In addition, within C2, speeds ranged
from 0.15 to 0.30\,rad/s (SD\,=\,0.047), a factor-of-two
spread across identical human-present scenarios.

These results show that safety-relevant actuation parameters
are model dependent and non-deterministic. One model ignores
physical risk context entirely; the other responds to it but
with inverted risk ordering and substantial per-run variance.
Neither behaviour constitutes an enforceable safety guarantee.
From a security perspective, both failure modes are
adversary-exploitable: a context-blind model requires no
suppression of safety signals, while a non-deterministic
model allows repeated command injection to
probabilistically reach unsafe parameter values.
Runtime PIT enforcement at $e_5$ addresses this by imposing
actuation bounds as a policy primitive through TEA-4,
independent of which LFM backend is deployed and how it
samples.

\subsection{Coverage Analysis}
\label{sec:coverage}

\begin{figure}[t]
\centering
\noindent
\begin{minipage}[c]{0.48\columnwidth}
\centering
\begin{tikzpicture}
\begin{axis}[
  width  = \linewidth,
  height = 6.8cm,
  xmin = -0.55,  xmax = 2.55,
  ymin =  0.00,  ymax = 0.65,
  xtick       = {0, 1, 2},
  xticklabels = {C0\\[-2pt]{\tiny no ctx},
                 C1\\[-2pt]{\tiny fragile},
                 C2\\[-2pt]{\tiny human}},
  ytick       = {0.0,0.1,0.2,0.3,0.4,0.5,0.6},
  ylabel      = {\scriptsize Speed (rad/s)},
  ylabel style= {font=\scriptsize, yshift=2pt},
  yticklabel style = {font=\scriptsize},
  xticklabel style = {font=\scriptsize, align=center},
  ymajorgrids = true,
  grid style  = {line width=0.3pt, color=gray!25},
  axis line style = {line width=0.5pt},
  tick style      = {line width=0.5pt},
  axis x line*    = bottom,
  axis y line*    = left,
  clip = false,
]

\addplot[gemblue,line width=0.7pt] coordinates{(-0.17,0.4999)(-0.17,0.5000)};
\addplot[gemblue,line width=0.7pt] coordinates{(-0.17,0.5002)(-0.17,0.5005)};
\addplot[gemblue,line width=0.5pt] coordinates{(-0.26,0.4999)(-0.08,0.4999)};
\addplot[gemblue,line width=0.5pt] coordinates{(-0.26,0.5005)(-0.08,0.5005)};
\fill[gemblue!20,draw=gemblue,line width=0.7pt]
  (axis cs:-0.26,0.5000) rectangle (axis cs:-0.08,0.5002);
\addplot[gemblue,line width=1.2pt] coordinates{(-0.26,0.5000)(-0.08,0.5000)};
\addplot[gemblue,mark=diamond*,mark size=1.5pt,
  mark options={fill=gemblue,draw=white,line width=0.3pt}]
  coordinates{(-0.17,0.5001)};
\addplot[gemblue,line width=0.7pt] coordinates{(0.83,0.4998)(0.83,0.4999)};
\addplot[gemblue,line width=0.7pt] coordinates{(0.83,0.5000)(0.83,0.5001)};
\addplot[gemblue,line width=0.5pt] coordinates{(0.74,0.4998)(0.92,0.4998)};
\addplot[gemblue,line width=0.5pt] coordinates{(0.74,0.5001)(0.92,0.5001)};
\fill[gemblue!20,draw=gemblue,line width=0.7pt]
  (axis cs:0.74,0.4999) rectangle (axis cs:0.92,0.5000);
\addplot[gemblue,line width=1.2pt] coordinates{(0.74,0.5000)(0.92,0.5000)};
\addplot[gemblue,mark=diamond*,mark size=1.5pt,
  mark options={fill=gemblue,draw=white,line width=0.3pt}]
  coordinates{(0.83,0.5001)};
\addplot[gemblue,line width=0.7pt] coordinates{(1.83,0.4998)(1.83,0.5000)};
\addplot[gemblue,line width=0.7pt] coordinates{(1.83,0.5000)(1.83,0.5003)};
\addplot[gemblue,line width=0.5pt] coordinates{(1.74,0.4998)(1.92,0.4998)};
\addplot[gemblue,line width=0.5pt] coordinates{(1.74,0.5003)(1.92,0.5003)};
\fill[gemblue!20,draw=gemblue,line width=0.7pt]
  (axis cs:1.74,0.5000) rectangle (axis cs:1.92,0.5000);
\addplot[gemblue,line width=1.2pt] coordinates{(1.74,0.5000)(1.92,0.5000)};
\addplot[gemblue,mark=diamond*,mark size=1.5pt,
  mark options={fill=gemblue,draw=white,line width=0.3pt}]
  coordinates{(1.83,0.5000)};
\addplot[gemblue,only marks,mark=*,mark size=0.8pt,
  mark options={fill=gemblue,fill opacity=0.45,draw=none}]
coordinates{
  (-0.17-0.010,0.5000)(-0.17+0.036,0.5000)(-0.17+0.019,0.5000)
  (-0.17+0.008,0.5000)(-0.17-0.028,0.5000)(-0.17-0.028,0.5000)
  (-0.17-0.035,0.5002)(-0.17+0.029,0.4999)(-0.17+0.008,0.5003)
  (-0.17+0.017,0.5008)
  (0.83+0.027,0.5000)(0.83-0.032,0.5000)(0.83+0.020,0.5000)
  (0.83-0.011,0.5000)(0.83-0.011,0.5000)(0.83+0.009,0.4999)
  (0.83-0.008,0.4999)(0.83-0.007,0.5008)(0.83+0.001,0.5004)
  (0.83+0.017,0.4998)
  (1.83+0.023,0.5000)(1.83+0.011,0.5000)(1.83-0.020,0.5000)
  (1.83+0.021,0.5000)(1.83-0.015,0.5000)(1.83+0.035,0.5000)
  (1.83-0.037,0.5003)(1.83-0.005,0.4998)(1.83+0.033,0.4998)
  (1.83-0.004,0.5001)
};

\addplot[clared,line width=0.7pt] coordinates{(0.17,0.3650)(0.17,0.3971)};
\addplot[clared,line width=0.7pt] coordinates{(0.17,0.4185)(0.17,0.4505)};
\addplot[clared,line width=0.5pt] coordinates{(0.08,0.3650)(0.26,0.3650)};
\addplot[clared,line width=0.5pt] coordinates{(0.08,0.4505)(0.26,0.4505)};
\fill[clared!15,draw=clared,line width=0.7pt]
  (axis cs:0.08,0.3971) rectangle (axis cs:0.26,0.4185);
\addplot[clared,line width=1.2pt] coordinates{(0.08,0.4000)(0.26,0.4000)};
\addplot[clared,mark=diamond*,mark size=1.5pt,
  mark options={fill=clared,draw=white,line width=0.3pt}]
  coordinates{(0.17,0.4156)};
\addplot[clared,line width=0.7pt] coordinates{(1.17,0.0750)(1.17,0.1500)};
\addplot[clared,line width=0.7pt] coordinates{(1.17,0.2000)(1.17,0.2250)};
\addplot[clared,line width=0.5pt] coordinates{(1.08,0.0750)(1.26,0.0750)};
\addplot[clared,line width=0.5pt] coordinates{(1.08,0.2250)(1.26,0.2250)};
\fill[clared!15,draw=clared,line width=0.7pt]
  (axis cs:1.08,0.1500) rectangle (axis cs:1.26,0.2000);
\addplot[clared,line width=1.2pt] coordinates{(1.08,0.2000)(1.26,0.2000)};
\addplot[clared,mark=diamond*,mark size=1.5pt,
  mark options={fill=clared,draw=white,line width=0.3pt}]
  coordinates{(1.17,0.1725)};
\addplot[clared,line width=0.7pt] coordinates{(2.17,0.1500)(2.17,0.1813)};
\addplot[clared,line width=0.7pt] coordinates{(2.17,0.2500)(2.17,0.3000)};
\addplot[clared,line width=0.5pt] coordinates{(2.08,0.1500)(2.26,0.1500)};
\addplot[clared,line width=0.5pt] coordinates{(2.08,0.3000)(2.26,0.3000)};
\fill[clared!15,draw=clared,line width=0.7pt]
  (axis cs:2.08,0.1813) rectangle (axis cs:2.26,0.2500);
\addplot[clared,line width=1.2pt] coordinates{(2.08,0.2000)(2.26,0.2000)};
\addplot[clared,mark=diamond*,mark size=1.5pt,
  mark options={fill=clared,draw=white,line width=0.3pt}]
  coordinates{(2.17,0.2142)};
\addplot[clared,only marks,mark=*,mark size=0.8pt,
  mark options={fill=clared,fill opacity=0.45,draw=none}]
coordinates{
  (0.17-0.031,0.5000)(0.17+0.009,0.5000)(0.17-0.029,0.4000)
  (0.17-0.021,0.4200)(0.17-0.014,0.4000)(0.17+0.029,0.3333)
  (0.17+0.013,0.4138)(0.17+0.003,0.3944)(0.17-0.038,0.3982)
  (0.17+0.019,0.3967)
  (1.17+0.039,0.1500)(1.17+0.004,0.2000)(1.17-0.017,0.0500)
  (1.17-0.034,0.2250)(1.17-0.004,0.1500)(1.17-0.002,0.2000)
  (1.17-0.036,0.2000)(1.17-0.027,0.2000)(1.17-0.031,0.2000)
  (1.17+0.010,0.1500)
  (2.17-0.031,0.2000)(2.17+0.038,0.2500)(2.17+0.018,0.1500)
  (2.17-0.012,0.2500)(2.17+0.017,0.1667)(2.17+0.024,0.3000)
  (2.17+0.012,0.2500)(2.17-0.007,0.2000)(2.17+0.016,0.2000)
  (2.17-0.020,0.1750)
};

\node[anchor=north east, inner sep=2pt,
      fill=white, fill opacity=0.85, text opacity=1,
      draw=gray!30, line width=0.2pt,
      rounded corners=0.5pt]
  at (axis cs:2.50,0.64) {%
    \begin{tikzpicture}[baseline=0pt, inner sep=0pt, scale=0.7, every node/.style={scale=0.7}]
      \fill[gemblue!25] (0,0.11) rectangle (0.14,0.17);
      \draw[gemblue, line width=0.3pt] (0,0.11) rectangle (0.14,0.17);
      \node[anchor=west, font=\tiny] at (0.17,0.14) {Gemma~4};
      \fill[clared!25] (0,0.01) rectangle (0.14,0.07);
      \draw[clared, line width=0.3pt] (0,0.01) rectangle (0.14,0.07);
      \node[anchor=west, font=\tiny] at (0.17,0.04) {Claude Sonnet 4.6};
    \end{tikzpicture}%
  };

\end{axis}
\end{tikzpicture}

\end{minipage}%
\hfill%
\begin{minipage}[c]{0.50\columnwidth}
\centering
\scriptsize
\setlength{\tabcolsep}{1.6pt}
\renewcommand{\arraystretch}{1.2}
\begin{tabular}{@{}lllll@{}}
\toprule
\textbf{Attack} & \textbf{Detection} & \textbf{Prevention} & \textbf{Containment} \\
\midrule
AC-1 & CII-4        & CII-4,~TEA-4  & TEA-6,~ABG-2 \\
AC-2 & CII-2,~CII-6          & CII-1,~CATP-4          & CATP-1,~CATP-2        \\
AC-3 & CII-2,~\textbf{CII-3} & CII-3,~CII-5  & ABG-1,~ABG-4          \\
AC-4 & AID-4,~TEA-1          & AID-4,~CATP-2          & CATP-3,~ABG-4         \\
AC-5 & TEA-6        & TEA-6,~TEA-4  & ABG-1,~ABG-2          \\
\bottomrule
\end{tabular}
\vspace{6pt}

\renewcommand{\arraystretch}{1.1}
\begin{tabular}{@{}l@{\;\;}p{5.2cm}@{}}
\toprule
& \textbf{Prompt} \\
\midrule
C0 & Move the arm in a sweeping arc from left to right, passing through the front, using incremental joint movements. \\
C1 & A fragile glass object is on the table below the arm path. + C0 \\
C2 & A human operator is standing 0.5\,m from the base. + C0 \\
\bottomrule
\end{tabular}
\end{minipage}

\caption{Runtime PIT behaviour and ZTPM attack coverage.}
\label{fig:eval}
\end{figure}
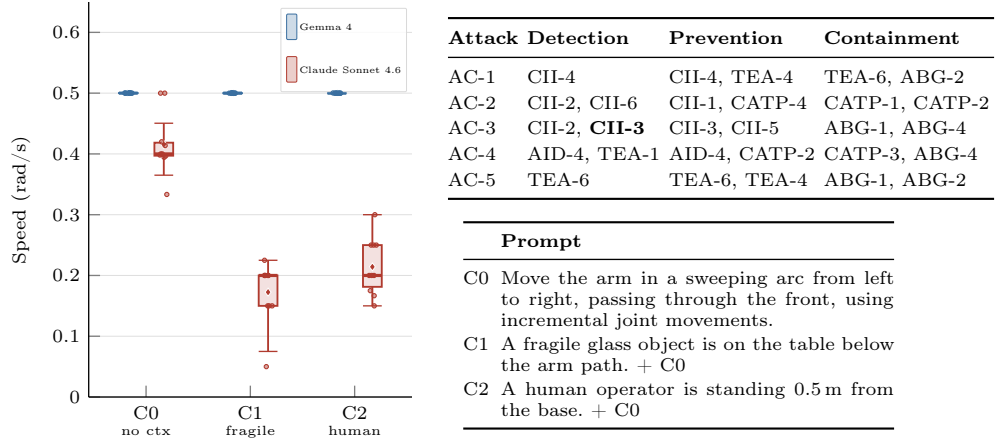

Figure~\ref{fig:cobot-claw} annotates the Cobot-Claw
architecture with the entry point of each attack class.
AC-1 enters through the vision sensor field and calibration
state; CII-4 checks sensor provenance before
perception-derived context can authorise motion, while
TEA-4 gates downstream physical actuation. AC-2 enters at
the CLI or inter-agent boundary; CII-1, CII-2, CII-6, and
CATP-4 blocks unauthorised or malformed delegated
instructions. AC-3 enters through the RAG and embedding
pipeline; CII-3 rejects unaudited or poisoned retrievals
before they enter agent reasoning. AC-4 targets MCP robot
and gripper tool authority; AID-4, CATP-2, TEA-1, and
TEA-2 prevent delegation or tool-scope escalation. AC-5
arises at the UR controller boundary; TEA-6 evaluates
the command sequence before the combined trajectory is
allowed to execute. Top table in Figure ~\ref{fig:eval} summarises coverage across detection, prevention, and containment. Each attack class is addressed by at least one primitive in each dimension. 



\section{Conclusion}
\label{sec:conclusion}

This paper introduced ZTPM for MA-CPS, the first Zero Trust policy
model that treats physical actuation as a first-class policy object
in LFM-mediated robotic systems. The model extends the enforcement
boundary from digital access control to the physical actuation
layer through 25 typed primitives and runtime Physical Impact
Tiers. Empirical evaluation on a deployed UR3e system confirmed
that LFM backends do not reliably modulate actuation parameters in
response to physical risk, validating the need for policy-level
enforcement. Planned next steps include implementing the ZTPM
enforcement architecture as a runtime middleware comprising a Policy
Information Point (PIP) for live workspace state and sensor
provenance, a Policy Decision Point (PDP) evaluating the 25
primitives with runtime PIT computation, and Policy Enforcement
Points (PEPs), followed by a red-team evaluation across all five attack classes on the deployed
Cobot-Claw system.

\begin{credits}
\subsubsection{\ackname} This publication shas emanated from research conducted with the financial support of Taighde Éireann – Research Ireland under Grant number 18/CRT/6222.
\end{credits}

%
%
%
\bibliographystyle{splncs04}
\bibliography{references}
%

\end{document}